\documentclass[12pt,letterpaper]{article}
\title{}
\usepackage{amsmath,amsfonts}
\usepackage{titling}
\usepackage{epsfig}
\usepackage{graphicx}
\usepackage{color}
\usepackage{csvsimple}
\usepackage[round]{natbib}
\usepackage{hyperref}
\usepackage{array}
\newcolumntype{?}{!{\vrule width 3pt}}

\DeclareMathOperator*{\argmin}{arg\,min}

\providecommand{\keywords}[1]
{
  \small	
  \textbf{\textit{Keywords---}} #1
}

\usepackage{authblk}
\usepackage{blindtext}

\title{Identifying group contributions in NBA lineups with spectral analysis}
\author{Stephen Devlin\thanks{Corresponding author, smdevlin@usfca.edu} }
\author{David Uminsky}
\affil{Department of Mathematics and Statistics, University of San Francisco}

\begin{document}
\maketitle

\begin{titlingpage}
\author{Stephen Devlin and David Uminsky}
    
    \begin{abstract}
We address the question of how to quantify the contributions of groups of players to team success. Our approach is based on spectral analysis, a technique from algebraic signal processing, which has several appealing features. First, our analysis decomposes the team success signal into components that are naturally understood as the contributions of player groups of a given size: individuals, pairs, triples, fours, and full five-player lineups. Secondly, the decomposition is orthogonal so that contributions of a player group can be thought of as pure: Contributions attributed to a group of three, for example, have been separated from the lower-order contributions of constituent pairs and individuals. We present detailed a spectral analysis using NBA play-by-play data and show how this can be a practical tool in understanding lineup composition and utilization.
   \end{abstract}


\end{titlingpage}


\section{Introduction}

A fundamental challenge in basketball performance evaluation is the team nature of the game. Contributions to team success occur in the context of a five-player lineup, and isolating the specific contribution of an individual is a difficult problem with a considerable history. Among the many approaches to the player evaluation problem are well-known metrics like player efficiency rating (PER), wins produced (WP), adjusted plus-minus (APM), box plus-minus (BPM), win shares (WS), value over replacement player (VORP), and offensive and defensive ratings (OR and DR) to name only a few \citep{BasketballReferenceGlossary}. While these individual player metrics help create a more complete understanding of player value, some contributions remain elusive. Setting good screens, ability to draw defenders, individual defense, and off-ball movement are all examples of important contributions that are difficult to measure and quantify. In part, these contributions are elusive because they often facilitate the success of a teammate who ultimately reaps the statistical benefit.  

Even beyond contributions that are difficult to quantify, the broader question of chemistry between players is a critical aspect of team success or failure. It is widely accepted that some groups of players work better together than others, creating synergistic lineups that transcend the sum of their individual parts. Indeed, finding (or fostering) these synergistic groups of players is fundamental to the role of a general manager or coach. There are, however, far fewer analytic approaches to identifying and quantifying these synergies between players. Such positive or negative effects among teammates represent an important, but much less well understood, aspect of team basketball.

In this paper we propose spectral analysis \citep{Diaconis:1988} as a novel approach to identifying and quantifying group effects in NBA play-by-play data. Spectral analysis is based on algebraic signal processing, a methodology that has garnered increasing attention from the machine learning community \citep{kakarala:2011, Kondor:2007, Kondor:2012}, and is particularly well suited to take advantage of the underlying structure of basketball data. The methodology can be understood as a generalization of traditional Fourier analysis, an approach whose centrality in a host of scientific and applied data analysis problems is well-known, and speaks to the promise of its application in new contexts from social choice to genetic epistasis and more \citep{Paudel:2013,Jurman:2008,Lawson:2006,Uminsky:2018,Uminsky:2019}. The premise of spectral analysis in a basketball context is simple: team success (appropriately measured) can be understood as a function on lineups. Such functions have rich structure which can be analyzed and exploited for data analytic insights.

Previous work in basketball analytics has addressed similar questions from a different perspective. Both \cite{kuehn2016accounting} and \cite{maymin2013nba} studied lineup synergies on the level of player skills. In \cite{maymin2013nba} the authors used a probabilistic framework for game events, along with simulated games to evaluate full-lineup synergies and find trades that could benefit both teams by creating a better fit on both sides. In \cite{kuehn2016accounting}, on the other hand, the author used a probabilistic model to determine complementary skill categories that suggest the effect of a player in the context of a specific lineup. Work in \cite{grassetti2019estimation} and \cite{grassetti2019play} modeled lineup and player effects in the Italian Basketball League (Serie A1) based on an adjusted plus-minus framework.  

Our approach is different in several respects. First, we study synergies on the level of specific player groups independent of particular skill sets. We also ignore individual production statistics and infer synergies directly from observed team success, as defined below. As a consequence of this approach, our analysis is roster constrained-- we don't suggest trades based on prospective synergies across teams. We can, however, suggest groupings of players that allow for more optimal lineups within the context of available players, a central problem in the course of an NBA game or season. Further, our approach uses orthogonality to distinguish between the contributions of a group and nested subgroups. So, for example, a group of three players that appears to exhibit positive synergies may, in fact, be benefiting from strong individual and pair contributions while the triple of players adds no particular value as a pure triple. We tease apart these higher-order correlations.

Furthermore, spectral analysis is not a model-based approach. As such, our methodology is notably free of modeling assumptions--rather than fitting the data, spectral analysis reports the observed data, albeit projected into a new basis with new information. Thus, it is a direct translation of what actually happened on the court (as we make precise below). As such, our methodology is at least complementary to existing work, and is also promising in presenting a new approach to understanding and appreciating the nuances of team basketball.

Finally, we note that while the methodology that underlies the spectral analysis approach is challenging, the resulting intuitions and insights are readily approachable. In what follows, we have stripped the mathematical details to a minimum and relegated them to references for the interested reader. The analysis, on the other hand, shows promise as a new and practical approach to a difficult problem in basketball analytics.     

\section{Data}
\label{Data}

We start with lineup level play-by-play data from the 2015-2016 NBA season. Such play-by-play data is publically available on ESPN.com or NBA.com, or can be purchased from websites like bigdataball.com, already processed into csv format. 
For a given team, we restrict attention to the 15 players on the roster having the most possessions played on the season, and filter the play-by-play data to periods of games involving only those players.  Next, we compute the aggregated raw plus-minus (PM) for each lineup. Suppose lineup $L$ plays against opposing lineup $M$ during a period of gameplay with no substitutions.  We compute the points scored by each lineup, as well as the number of possessions for both lineups during that stretch of play. For example, if lineup $L$ scored 6 points in 3 possessions and lineup $M$ scored 3 points in 2 possessions, then their plus-minus is computed as the difference in points-per-possession times possessions. Thus, for $L$ the plus-minus is $(\frac{6}{3} - \frac{3}{2})3 = 1.5$ while for $M$ the plus-minus is 
$(\frac{3}{2} - \frac{6}{3})2 = -1$. Summing over all of lineup $L$'s possessions gives the total aggregate plus-minus for lineup $L$ which we denote by $\text{pm}_L$. 

Since a lineup consists of 5 players on the floor, there are $3003={15\choose 5}$ possible lineups, though most see little or no playing time. We thus naturally arrive at a function on lineups by associating with $L$ the value of that lineup's aggregate plus-minus, and write $f(L)=\text{pm}_L$. We call $f$ the team success function. This particular success metric has the advantage of being simple and intuitive. Moreover, by summing over all lineups we recover the value of the team's cumulative plus-minus, which is highly correlated with winning percentage. The function $f$ will serve as the foundation for our analysis, but we note that for what follows, any quantitative measure of a lineup's success could be substituted in its place.

\section{Methodology}
\label{methodology}

Our goal is now to decompose the function $f$ in a way that sheds light on the various group contributions to team success. The groups of interest are generalized lineups, meaning groups of all sizes, from individual players to pairs, triples, groups of four, and full five-player lineups. Our primary tool is spectral analysis, which uses the language of representation theory \citep{serre2012linear} to understand functions on lineups.

Observe that a full lineup is an unordered set of five players. Any reshuffling of the five players on the floor, or the ten on the bench, does not change the lineup under consideration. Moreover, given a particular lineup, a permutation (or reshuffling) of the fifteen players on the team will result in a new lineup. The set of such permutations has a rich structure as a mathematical group. In this case, all possible permutations of fifteen players are described by $S_{15}$: the symmetric group on 15 items \citep{Dummit:2004}. Furthermore, the set $X$ of five-player lineups naturally reflects this group structure (as a homogeneous space). Most importantly for our purposes, the set of functions on lineups has robust structure with respect to the natural action of permutations on functions. This structure is well understood and can be exploited for data analytic insights as we show below. By way of analogy, just as traditional Fourier analysis looks to decompose a time series into periodicities that can reveal a hidden structure (weekly or seasonal trends, say), our decomposition of $f$ will reveal group effects in lineup-level data. 

Let  $L(X)$ denote the collection of all real valued functions on five-player lineups. This set is a vector space with the usual notions of sum of functions, multiplication by scalars, and an inner product given by 
\begin{equation}
\label{inner product}
\langle g,h \rangle =\frac{1}{|X|}\sum_{x\in X} g(x)h(x).
\end{equation}
The dimension of $L(X)$ is equal to the number of lineups, $3003={15\choose 5}$. In light of the permutation group's action on $L(X)$ as mentioned above,  $L(X)$ admits a natural (invariant and irreducible) decomposition as follows:
\begin{equation}
\label{decomposition}
L(X)=V_0\oplus V_1 \oplus V_2\oplus V_3\oplus V_4\oplus V_5 .
\end{equation}
Each $V_i$, with $0\le i \le 5$ is a vector subspace with data analytic significance. Rather than give a self contained treatment of this decomposition, we refer to \cite{Diaconis:1988} and \cite{Dummit:2004}, and here, simply note that each space is spanned by the matrix coefficients of the irreducible representations of the group $S_{15}$ associated with Young tableaux of shape $(10,5)$. We can gain some intuition for the decomposition by considering the lower-order spaces as follows. An explicit computation of the decomposition is given in section \ref{sec:toy example section} below for a toy example.  

Take $\delta_L$ to be the indicator function of a fixed lineup $L$, so that $\delta_L(L)=1$, while $\delta_{L}(L')=0$ for any other lineup $L'$. As above, $X$ is the set of all possible lineups, and 
\begin{equation}
\label{meanspace}
\delta=\sum_{L\in X}\delta_L.
\end{equation}
If we act on the function $\delta$ by reshuffling lineups (this is the action of the permutation group $S_{15}$), we see that while the terms in the summation in (\ref{meanspace}) get reordered, the function itself remains unchanged. (See section \ref{sec:toy example section} below for details.) Thus, the one-dimensional space spanned by $\delta$ is invariant under lineup reshuffling and represents the mean value of the function $f$ since we can write $f=c\delta+(f-c\delta)$. Here, $c$ is just the average value of $f$ and $c\delta$ is the best possible constant approximation to $f$. The function $f-c\delta$ represents the original data, but now centered with mean zero, and orthogonal to the space of constant functions with respect to the inner product in (\ref{inner product}). The space spanned by $\delta$ is $V_0$ in (\ref{decomposition}). 

To understand $V_1$, we start with indicator functions for individual players. Given a player $i$, define $\delta_i=\sum_{L\in\mathcal{L}_i}\delta_{L}-m\delta$ where the sum is over all lineups that include player $i$ and four other players, and $m$ is a constant chosen so that $\delta_i$ is orthogonal to $\delta$. One can show that the space spanned by $\{\delta_1,\delta_2,\ldots\delta_{15}\}$ is again stable under lineup reshuffling. (Though the set of individual indicator functions is linearly dependent, and only spans a 14-dimensional space as we'll see below.) 

The decomposition continues in an analogous way, though the computations become more involved.  Several computational approaches are described in \cite{Diaconis:1988} and \cite{maslen:2003}. In our case of the symmetric group $S_{15}$ acting on lineups, we employ the method in \cite{maslen:2003}, which involves first computing the adjacency matrix of an associated {\it Johnson graph} $J(15,5)$. It turns out that $J(15,5)$ has 6 eigenvalues, each of which is associated with one of the effect spaces: zero (mean), and first through fifth-order spaces. Specifically, the largest eigenvalue is simple and is associated with the one-dimensional mean space; the second largest eigenvalue is associated with the first-order space, etc. It is now a matter of computing an eigenbasis for each space, and using it to project the data vector onto each eigenspace to give the orthogonal decomposition used in (\ref{decomposition}). It is also worth noting that spectral analysis includes the traditional analysis of variance as a special case, a connection suggested by the discussion above and further explained in \cite{Diaconis:1988}.

The decomposition in (\ref{decomposition}) is particularly useful for two reasons. First, each $V_i$ can be interpreted as the space of functions encoding $i$-th order effects. For instance, one can see that $V_1$ is naturally understood as encoding first-order individual effects beyond the mean. Thus, the projection of $f$ onto $V_1$  can be understood as that part of team success $f$ attributable to the contributions of individual players. Similarly $V_2$ includes effects attributable to pure player pairs (individual contributions have been removed), and the corresponding projection of $f$ in $V_2$ gives the contributions of those pairs to team success. $V_3$ encodes contributions of groups of three, and so on. These interpretations follow from the fact that each subspace in the decomposition of $L(X)$ is invariant under the natural reshuffling action of $S_{15}$ on lineups. It is also worth noticing that the lineup success function is completely recovered via its projections onto the order subspaces in (\ref{decomposition}). If we write $f_i$ for the projection of $f$ onto $V_i$, then $f=f_0+f_1+f_2+f_3+f_4+f_5$. As such, the spectral decomposition gives a complete description of the original data set with respect to a new basis grounded in group contributions.  

Secondly, the decomposition in (\ref{decomposition}) is orthogonal (signified by the $\oplus$ notation). From a data analytic perspective, this means that there is no overlap among the spaces, and group effects are independent. Thus, for instance, a contribution attributed to a group of three players can be understood as a pure third-order contribution. All constituent pair and individual contributions have been removed and quantified separately in the appropriate lower-order spaces. We thus avoid erroneous attribution of success due to multicollinearity among groups. For example, is a big three really adding value as a triple, or is its success better understood as a strong pair plus an individual? The spectral decomposition in (\ref{decomposition}) provides a quantitative basis for answering such questions.  

The advantage of the orthogonality of the spaces in (\ref{decomposition}), however, presents a challenge with respect to direct interpretation of contributions for particular groups. This is evident when considering the dimension of each of the respective effect spaces in Table \ref{ProjSpaceDimensions}, which is strictly smaller than the number of groups of that size we might wish to analyze.

\begin{table}
\centering
\begin{tabular}{ccc}
\hline
Space & Dimension & Number of Groups\\ 
\hline
$V_0$ & 1 & --\\ 
$V_1$ & 14 & 15\\  
$V_2$ &  90 & 105\\  
$V_3$ &  350 &455\\ 
$V_4$ & 910 & 1365\\  
$V_5$ & 1638 & 3003\\ 
\hline 
\end{tabular}
\caption{Dimension of each effect space, along with the number of natural groups of each size.}
\label{ProjSpaceDimensions}
\end{table} 
Since we have rosters of fifteen players, there are fifteen individual contributions to consider. The space $V_1$, however, is 14-dimensional. Similarly, while $V_2$ includes all of the contributions to $f$ attributable to pairs of players, it does so in a 90-dimensional space despite the fact that there are $105={15\choose 2}$ natural pairs of players to consider. The third-order space $V_3$ has dimension 350 while there are 455 player triples, and so on. 

We deal with this issue using Mallows' method of following easily interpretable vectors as in \cite{Diaconis:1988}. Let $g$ be a group of players. For example, if players are labeled 1 through 15, then a particular triple might be $g=\{1,2,7\}$. Let $\phi_g$ be the indicator function associated with $g$, i.e., the function that takes the value 1 when all three players 1, 2, and 7 are in a lineup, and outputs 0 otherwise. The function $\phi_g$ is intuitively associated with the success of the group $g$ (though it is not invariant under reshuffling and is not orthogonal to nested lower-order groups). 

To quantify the contribution of $g$ (as a pure triple) to the success of the team as measured by $f$, project both $\phi_g$ and $f$ onto $V_3$ and take the inner product of the projections: $\langle pr_{V_3}(\phi_g), pr_{V_3}(f)\rangle = \langle pr_{V_3}(\phi_g), f_3\rangle$. After projecting onto $V_3$ we are left with only the third-order components of $\phi_g$ and $f$. The resulting inner product is a weighted cosine similarity that indicates the extent to which the pure triple $g$ is correlated with the team's success $f$. Larger values of this inner product reflect a stronger synergy between the triple of players $\{1,2,7\}$, while a negative value indicates that, after removing the contributions of the constituent individuals and pairs, spectral analysis finds this particular group of three ineffective. In the results below we show how this information might be useful in evaluating lineups.

\section{Two-On-Two Basketball}
\label{sec:toy example section}
To ground the ideas of the previous section we present a small-scale example in detail. Consider a version of basketball where a team consists of 5 players, two of which play at any given moment. The set of possible lineups consists of the ten unordered pairs $\{i,j\}$ with $i,j\in\{1,2,3,4,5\}$ and $i\ne j$. The symmetric group $S_5$ acts on lineups by relabeling, and we extend this action to functions on lineups as follows. Given a permutation $\pi$, a function $h$, and a lineup $L$, define    
\begin{equation}
(\pi\cdot h)(L)=h(\pi^{-1}L). 
\end{equation}
Therefore, if $\pi$ is the permutation $(123)$, taking player 1 to player 2, player 2 to player 3, player 3 to player 1, and leaving everyone else fixed, and if $L$ is the lineup $\{1,3\}$, then
\begin{equation}
(\pi\cdot h)(L) = h(\pi^{-1}\{1,3\}) = h(\{3,2\}).
\end{equation}
The use of the inverse is necessary to ensure that the action on functions respects the operation in the group, that is, so that $(\tau\pi)\cdot h = \tau\cdot (\pi\cdot h)$ \citep{Dummit:2004}.

Following a season of play, we obtain a success function that gives the plus-minus (or other success metric) of each lineup. We might observe a function like that in Table \ref{ToyLineupFunction}.

\begin{table}[ht]
\centering
\begin{tabular}{cc|cc}
\hline
  $L$ & $f(L)$ & $L$ & $f(L)$\\ \hline
  $\{1,2\}$ &22  &$\{2,4\}$& 35\\
  $\{1,3\}$ &18  &$\{2,5\}$& 26\\
  $\{1,4\}$ &3  &$\{3,4\}$& 84\\
  $\{1,5\}$ &58  &$\{3,5\}$& 25\\
  $\{2,3\}$ &93  &$\{4,5\}$& 2\\
\end{tabular}
\caption{Success function for two-player lineups.}
\label{ToyLineupFunction}
\end{table}

Summing $f(L)$ over all lineups that include a particular player gives individual raw plus-minus as in Table \ref{ToyIndPM}.
\begin{table}[ht]
\centering
\begin{tabular}{ccc}
\hline
  Player & PM &Rank  \\ \hline
  1 &  101&5 \\
  2 &  176&2\\
  3 &  220&1\\
  4 &  124&3\\
  5 &  111& 4\\
\end{tabular}
\caption{Preliminary analysis of sample team using individual plus-minus (PM), which is the sum of the lineup PM over lineups that include a given individual.}
\label{ToyIndPM}
\end{table}
Player 3 is the top rated individual, followed by 2, 4, 5, and 1. 
Lineup rankings are given by $f(L)$ itself, which shows $\{2,3\},\{3,4\}$, and $\{1,5\}$ as the top three. 

Now compare the analysis above with spectral analysis. In this context the vector space of functions on lineups is 10-dimensional and has a basis consisting of vectors $\delta_{\{i,j\}}$ that assign the value 1 to lineup $\{i,j\}$ and 0 to all other lineups. The decomposition in (\ref{decomposition}) becomes
\begin{equation}
\label{toy decomposition}
V=V_0\oplus V_1\oplus V_2.
\end{equation}
Define $\delta = \sum_{\{i,j\}}\delta_{\{i,j\}}$. The span of $\delta$ is the one-dimensional subspace $V_0$ of constant functions. Moreover, $V_0$ is $S_5$ invariant since for any relabeling of players given by $\pi$, we have $\pi\cdot\delta =\delta$.
Given a function $f$ in $V$, its projection $f_0$ on $V_0$ will assigns to each lineup the average value of $f$, in this case 36.6. 


First order (or individual) effects beyond the mean are in encoded in $V_1$. Explicitly, define $\delta_1 = \sum_i\delta_{\{1,i\}}-\frac{2}{5}\delta$, with $\delta_2,\delta_3,$ and $\delta_4$ defined analogously. One can check that the 4-dimensional vector space spanned by $\{\delta_1,\delta_2,\delta_3,\delta_4\}$, is $S_5$ invariant, and is orthogonal to $V_0$. Since the mean has been subtracted out and accounted for in $V_0$, a vector in $V_1$ represents a pure first order effect. Note that $\delta_5(x)=\sum_i\delta_{\{5,i\}}-\frac{2}{5}\delta$ can be written $\delta_5=-\delta_1-\delta_2-\delta_3-\delta_4$. Consequently, $V_1$ is 4-dimensional even though there are five natural first order effects to consider: one for each player.

Finally, the orthogonal complement of $V_0\bigoplus V_1$ is the 5-dimensional $S_5$ invariant subspace $V_2$. $V_2$ gives the contribution to $f$ from {\it pure pairs}, or pure second order effects after the mean and individual contributions are removed. 
The three subspaces $V_0$, $V_1$, and $V_2$ are all irreducible since none contains a nontrivial $S_5$ invariant subspace.


We can now project $f$ onto $V_0, V_1$, and $V_2$. All together we have $f=f_0+f_1+f_2$:
\begin{equation}
\label{toy_decomposition}
f\left(
\begin{array}{c}
\{1,2\}\\
\{1,3\}\\
\{1,4\}\\
\{1,5\}\\
\{2,3\}\\
\{2,4\}\\
\{2,5\}\\
\{3,4\}\\
\{3,5\}\\
\{4,5\}\\
\end{array}\right)
=\left[ \begin{array}{r}
                      22 \\
                         18 \\
                         3 \\
                         58 \\
                         93\\
                         35 \\
                         26 \\
                         84 \\
                         25 \\
                         2 \\
                       \end{array}
                       \right]
=
          \left[ \begin{array}{c}
                      36.6 \\
                         36.6 \\
                         36.6 \\
                         36.6 \\
                         36.6\\
                         36.6 \\
                         36.6 \\
                         36.6 \\
                         36.6 \\
                         36.6 \\
                       \end{array}
                       \right]
+
\left[ \begin{array}{r}
                         -5.27 \\
                         9.40 \\
                         -22.60 \\
                         -26.93 \\
                         34.40\\
                         2.40 \\
                         -1.93 \\
                         17.07 \\
                         12.73 \\
                         -19.27 \\
                       \end{array}\right]
+
\left[ \begin{array}{r}
                         -9.33 \\
                         -28.00 \\
                         -11.00 \\
                         48.33 \\
                         22.00\\
                         -4.00 \\
                         -8.67 \\
                         30.33 \\
                         -24.33 \\
                         -15.33 \\
                       \end{array}\right]
\end{equation}


Turning to the question of interpretability, section \ref{methodology} proposes Mallows' method of using readily interpretable vectors projected into the appropriate effect space.  
To that end, the individual indicator function $\phi_{\{2\}}=\delta_{\{1,2\}}+\delta_{\{2,3\}}+\delta_{\{2,4\}}+\delta_{\{2,5\}}$ is naturally associated with player 2: $\phi_{\{2\}}(L)=1$ when player 2 is in $L$ and is 0 otherwise.  We quantify the effect of player $2$ by projecting $\phi_{\{2\}}$ and $f$ into $V_1$, and then taking the dot product of the projections. For a lineup like $\{2,3\}$, we take the dot product of the projections of the lineup indicator function $\delta_{\{2,3\}}$, and $f$, in $V_2$.   
Note that player 2's raw plus-minus is the inner product of $10\cdot f$ with the interpretable function $\phi_{\{2\}}$. Similarly $f(\{i,j\})$ is $10\cdot\langle f,\phi_{\{i,j\}}\rangle$. The key difference is that spectral analysis uses Mallow's Method {\it after} projecting onto the orthogonal subspaces in (\ref{toy decomposition}).


Contributions from spectral analysis as measured by Mallows' method are given in Table \ref{toyspec} for both individuals and (two-player) lineups.
\begin{table}[ht]
{\footnotesize
\centering
\label{toyspec}
\begin{tabular}{cc?cccc|cccc}
\hline
  Individual& Spec&Pair &Spec &Rank &$f$ Rank& Pair & Spec &Rank & $f$ Rank\\ \hline
 \{1\}&-45.4& \{1,2\} &-9.3  & 6 &7&\{2,4\}&-4&4&4\\
 \{2\}&29.6&\{1,3\} &-28  &10 &8&\{2,5\} &-8.7&5&5\\
 \{3\}&73.6&\{1,4\}& -11 & 7 &9&\{3,4\}& 30.3&2&2\\
 \{4\}&-22.4&\{1,5\}&48.3 & 1 &3&\{3,5\}& -24.3&9&6\\
 \{5\}&-35.4&\{2,3\}& 22 &3& 1&\{4,5\} & -24&8&10\\
\end{tabular}
\caption{Spectral value (Spec) for each individual player and two-player lineup, and rank of each lineup, along with the preliminary rank given by $f$.}
}
\end{table}
The table also includes both the spectral and preliminary (based on $f$) rankings of each lineup. Note that lineup $\{2,3\}$ drops from the best pair to the third best pure pair. Once we account for the contributions of players two and three as individuals, the lineup is not nearly as strong as it appears in the preliminary analysis. We find stronger pair effects from lineups $\{1,5\}$ and $\{3,4\}$. All remaining lineups are essentially ineffective in that their success can be attributed to the success of the constituent individuals rather than the pairing. Interesting questions immediately arise. What aspects of player four's game result in a more effective pairing with player three, the team's star individual player, than the pairing of three with two, the team's second best individual?   What is behind the success of the $\{1,5\}$ lineup? 
These considerations are relevant to team construction, personnel considerations, and substitution patterns. We pursue this type of analysis further in the context of an actual NBA team below.
 
{}

\section{Results and Discussion}

A challenge inherent in working with real lineup-level data is the wide disparity in the number of possessions that lineups play. Most teams have a dominant starting lineup that plays far more possessions than any other. For example, the starting lineup of the '16 Golden State Warriors played approximately 1140 possessions while the next most used lineup played 535 possessions. Only 12 lineups played more than 100 possessions for the Warriors on the season. For the Boston Celtics, the starters played 1413 possessions compared to 257 for the next most utilized, with 13 lineups playing more than 100 possessions. By contrast, the Celtics had 255 lineups that played fewer than 10 possessions (but at least one), and the Warriors had 236. Numbers are similar across the league. This is another reason for using raw plus-minus in defining the team success function $f$ on lineups. A metric like per-possession lineup plus-minus breaks down in the face of large numbers of very low possession lineups and a few high possession lineups. Still, we want to identify potentially undervalued and underutilized groups of players-- especially for smaller groups like pairs and triples where there are many more groups that do play significant numbers of possessions. Another consideration is that over time, lineups with large numbers of possessions will settle closer to their true mean value while lineups with few possessions will be inherently noisier. As a result, we perform the spectral analysis on $f$ as described in section \ref{methodology} above, and then normalize the spectral contribution by the log of possessions played by each group. We call the result {\it spectral contribution per log possession} (SCLP). This balances the considerations above and allows strong lower possession groups to emerge while not over-penalizing groups that do play many possessions. 

Despite these challenges, however, we'll see below that there are significant insights to be gained in working with lineup level data. Moreover, since spectral analysis is a non-model-based description of complete lineup-level game data, it has the advantage of maintaining close proximity to the actual gameplay observed by coaches, players, and fans. There are always five players on the floor, so all data begins at the level of full lineups.

Consider the first order effects for the 15-16 Golden State Warriors in Table \ref{GSWFirstTable}. Draymond Green, Stephen Curry, and Klay Thompson are the top three players. The ordering, specifically Green ranked above Curry, is perhaps interesting, though it's worth noting that this ordering agrees with ESPN's real plus-minus (RPM). (Green led the entire league in RPM in 15-16.) Other metrics like box plus-minus (BPM) and wins-above-replacement (WAR) rank Curry higher. Because SCLP is based on ability of lineups to outscore opponents when the player is on the floor (like RPM), however, as opposed to metrics like BPM and WAR which are more focused on points produced, the ordering is defensible. 

\begin{table}[ht]
\begin{center}
\begin{tabular}{lccc}
Player & SCLP  & PM & Poss\\
\hline
Draymond Green&       17.2&  1038.4 & 5800\\
Stephen Curry   &    15.9 &  978.7&  5610\\
Klay Thompson    &   12.0 &  808.6&  5453\\
Andre Iguodala     &   3.5 &  436.1 & 3516\\
Andrew Bogut    &    2.8 &  403.6 & 2951\\
\hline\hline
Marreese Speights &      -7.4 &   20.0&  1630\\
Ian Clark   &    -9.8 &  -51.9&  1108\\
Anderson Varejao &     -11.1 &  -34.4&   368\\
Jason Thompson    &  -11.2&   -33.8 &  339\\
James Michael McAdoo &     -12.1&   -85.0&   526\\
\end{tabular}
\caption{Top and bottom five first-order effects for GSW. SCLP is the spectral contribution per log possession, PM is the player's raw plus-minus, and Poss is the number of possessions for that player.}\label{GSWFirstTable}
\end{center}
\end{table}

In fact, a closer look at the interpretable vector $\phi_i$ associated with individual player $i$ (as described in sections \ref{methodology} and \ref{toyspec}) reveals that $\phi_i=\delta_i+c\cdot \delta$, so is just a non-mean-centered  version of the first order invariant functions that span $V_1$. Consequently, the spectral contribution (non-possession normalized) is a linear function of individual plus-minus, so reflects precisely that ordering. This is not the case for higher-order groups, however, which is where we focus the bulk of our analysis. 

The second-order effects are given in in Table \ref{GSWSecondTable}, and quantify the contributions of player pairs, having removed the mean, individual, and higher-order group effects. The top and bottom five pairs (in terms of SCLP) are presented here, with more complete data in Table \ref{AGSWSecondTable} in the appendix. 
\begin{table}[ht]
{\footnotesize
\begin{center}
\begin{tabular}{llccc}
\hline 
P1 &P2 & SCLP  & PM & Poss\\
\hline
Draymond Green &  Stephen Curry    &   13.3&  979.9 & 5102\\
Stephen Curry      &   Klay Thompson  &     11.2&  827.8 & 4311\\
Draymond Green  &  Klay Thompson   &    11.1&  847.8 & 4678\\
Leandro Barbosa  &  Marreese Speights&        5.3&   76.2  & 983\\
Draymond Green  &  Andre Iguodala&        4.3&  490.0 & 2165\\
\hline\hline           
Draymond Green   &             Ian Clark    &   -7.2 &  33.3  & 424  \\                        
Klay Thompson      &             Leandro Barbosa &      -7.2 &   4.8   &349 \\                 
Stephen Curry        &             Ian Clark    &   -8.1  & 14.0 &  220  \\                    
Draymond Green   &            Anderson Varejao &      -9.5  &  7.2  & 217\\                    
Stephen Curry        &            Anderson Varejao  &    -10.1 & -26.9  & 237 \\  \hline     
\end{tabular}
\caption{Top and bottom five SCLP pairs with at least 200 possessions, along with raw plus-minus and possessions.}\label{GSWSecondTable}
\end{center}
}
\end{table}
Even after accounting for and removing their strong individual contributions, however, it is notable that Green--Curry, Curry--Thompson, and Green--Thompson are the dominant pair contributors by a considerable margin, with SCLP values that are all more than twice as large as for the next largest pair (Barbosa--Speights). These large positive SCLP values represent true synergies: These pairs contribute to team success {\it as pure pairs}. The fact that the individual contributions of the constituent players are also positive results in a stacking of value within a lineup that provides a quantifiable way of assessing whether the whole does indeed add to more than the sum of its parts. 

Reserves Leandro Barbosa, Mareese Speights, and Ian Clark, on the other hand, were poor individual contributors, but manage to combine effectively in several pairs. In particular, the Barbosa--Speights pairing is notable as the fourth best pure pair on the team (in 983 possessions). After accounting for individual contributions, lineups that include the Barbosa--Speights pairing benefited from a real synergy that positively contributed to team success. This suggests favoring, when feasible, lineup combinations with those two players together to leverage this synergy and mitigate their individual weaknesses.

Tables \ref{SmallBogutPairs} and \ref{SmallLivingstonPairs} show pair values for players Andrew Bogut and Shaun Livingston (again in pairs with at least 150 possessions, and with more detailed tables in the appendix). Both players are interesting with respect to second order effects. While Bogut was a positive individual contributor, and was a member of the Warriors' dominant starting lineup that season, he largely fails to find strong pairings. His best pairings are with Klay Thompson and Harrison Barnes, while he pairs particularly poorly with Andre Iguodala (in a considerable 785 possessions). This raises interesting questions as to why Bogut's style of play is better suited to players like Thompson or Barnes rather than players like Curry or Iguodala. Also noteworthy is the fact that the Bogut--Iguodala pairing has a positive plus-minus value of 107. The spectral interpretation is that this pairing's success should be attributed to the individual contributions of the players, and once those contributions are removed, the group lacks value as a pure pair.     

\begin{table}[h!]
{\footnotesize
\begin{center}
\begin{tabular}{llccc}
\hline 
P1 &P2 & SCLP  & PM & Poss\\
\hline
Andrew Bogut  &  Klay Thompson& 3.7 &      394.3&             2637\\ 
Andrew Bogut &   Harrison Barnes &   2.1 &   206.2&           1527 \\ 
Andrew Bogut &   Stephen Curry &    1.6 &    378.5&            2530 \\  \hline
Andrew Bogut  &  Andre Iguodala &   -2.1 &      107.0 &          785 \\ 
\hline
\end{tabular}
\caption{Select pairs involving Andrew Bogut (with at least 150 possessions).}\label{SmallBogutPairs}
\end{center}
}
\end{table}

\begin{table}[h!]
{\footnotesize
\begin{center}
\begin{tabular}{llccc}
\hline 
P1 &P2 & SCLP  & PM &Poss\\
\hline
Shaun Livingston   &   Anderson Varejao  &      2.0  & -1.5   & 174    \\
  Shaun Livingston   &  Marreese Speights  &      1.6 &  17.8  & 1014  \\
 Shaun Livingston    &   Draymond Green    &   1.2 & 323.6  & 1486\\ \hline
   Shaun Livingston    &  Andre Iguodala      &  -1.3  & 65.2  & 1605          \\
  Shaun Livingston     &    Klay Thompson   &   -3.6 & 111.8  & 1412        \\
\hline
\end{tabular}
\caption{Select pairs involving Shaun Livingston (with at least 150 possessions).}\label{SmallLivingstonPairs}
\end{center}
}
\end{table}

Shaun Livingston, on the other hand, played an important role as a reserve point guard for the Warriors. Interestingly, Livingston's worst pairing by far was with Klay Thompson. Again, considering the particular styles of these players compels interesting questions from the perspective of analyzing team and lineup compositions and playing style. It's also noteworthy that this particular pairing saw 1412 possessions, and it seems entirely plausible that its underlying weakness was overlooked due to the healthy 111.8 plus-minus with that pair on the floor. The success of those lineups should be attributed to other, better synergies. For example, one rotation added Livingston as a sub for Barnes (112 possessions). Another put Livingston and Speights with Thompson, Barnes, and Iguodala (70 possessions).  
Finally, it's also interesting to note that Livingston appears to pair better with other reserves than with starters (save Draymond Green,  further highlighting Green's overall value), an observation that raises important questions about how players understand and occupy particular roles on the team. 

Table \ref{GSWThirdTable} shows the best and worst triples with at least 200 possessions. 
\begin{table}
{\footnotesize
\begin{center}
\begin{tabular}{lllrrr}
\hline 
P1 & P2 &P3 & SCLP & PM & Poss \\ \hline
Draymond Green &   Stephen Curry  &    Klay Thompson    &   12.6 & 812.7   &    4085  \\
Draymond Green   & Klay Thompson  & Harrison Barnes  &      5.9 & 427.3 &  2473 \\
Draymond Green  &  Stephen Curry  &   Andre Iguodala    &    5.8 & 464.8  & 1830 \\
Stephen Curry  &  Klay Thompson   & Harrison Barnes   &     5.7 & 416.5  &  2431 \\
Stephen Curry &   Klay Thompson   &    Andrew Bogut    &    4.9&  382.2 &   2296 \\
\hline\hline
Stephen Curry & Andre Iguodala & Brandon Rush & -3.8  & -13.5 & 207 \\
Draymond Green & Stephen Curry & Marreese Speights & -4.1  & 97.9  & 299 \\
Draymond Green & Klay Thompson & Marreese Speights & -4.5  & 52.2  & 250 \\
Draymond Green & Klay Thompson & Ian Clark & -5.8  & 9.8   & 316 \\
Draymond Green & Stephen Curry & Ian Clark & -7.4  & 14.5  & 205 \\
\hline
\end{tabular} 
\caption{Best and worst third-order effects for GSW with at least 200 possessions.}\label{GSWThirdTable}
\end{center}
}
\end{table}
The grouping of Green--Curry--Thompson is far and away the most dominant triple, and safely (and unsurprisingly) earns designation as the Warriors' big three. Other notable triples include starters like Green and Curry or Green and Thompson together with Andre Iguodala who came off the bench, and more lightly used triples like Curry--Barbosa--Speights who had an SCLP of 4.6 in 245 possessions. Analyzing subpairs of these groups shows a better stacking of synergies in the triples that include Iguodala--he pairs well with Green, Curry, and Thompson in the second order space as well, while either of Barbosa or Speights paired poorly with Curry. Still, Barbosa with Speights was quite strong as a pair, and we see that the addition of Curry does provide added value as a pure triple. Interesting ineffective triples include Iguodala and Bogut with either of Curry or Green, especially in light of the fact that Bogut--Iguodala  was also a weak pairing (see detailed tables in the appendix).

Figure \ref{GSW3scatter} shows that the most effective player-triples as identified by spectral analysis are positively correlated with higher values of plus-minus.  
\begin{figure}[ht]
\centering
\includegraphics[width = \textwidth]{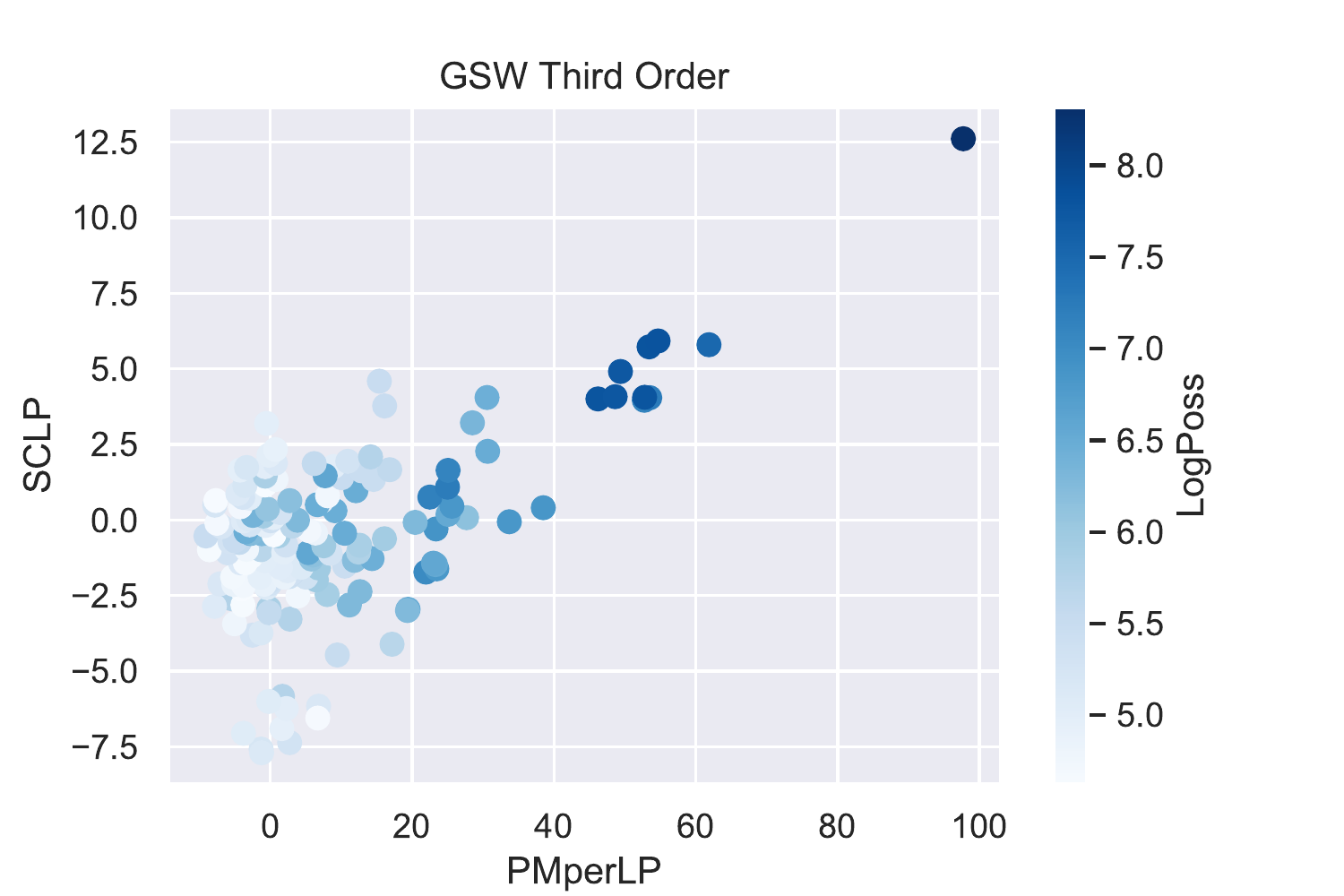}
\caption{Third-order effects for triples with more than 100 possessions the 2015-2016 Golden State Warriors. The $x$-axis gives the group's plus-minus per log possession (PMperLP) while the $y$-axis shows the spectral contribution per log possession (SCLP). Observations are shaded by number of possessions.}
\label{GSW3scatter}
\end{figure}
As raw group plus-minus decreases, however, we see considerable variation in the spectral contributions of the groups (and in number of possessions played). This suggests the following narrative: while it may be relatively easy to identify the team's top groups, it is considerably more difficult to identify positive and negative synergies among the remaining groups, especially when controlling for lower-order contributions. Spectral analysis suggests several opportunities for constructing more optimal lineups with potential for untapped competitive advantage, especially when more obvious dominant groupings are unavailable. 

\begin{figure}[h!]
\centering
\includegraphics[width = \textwidth]{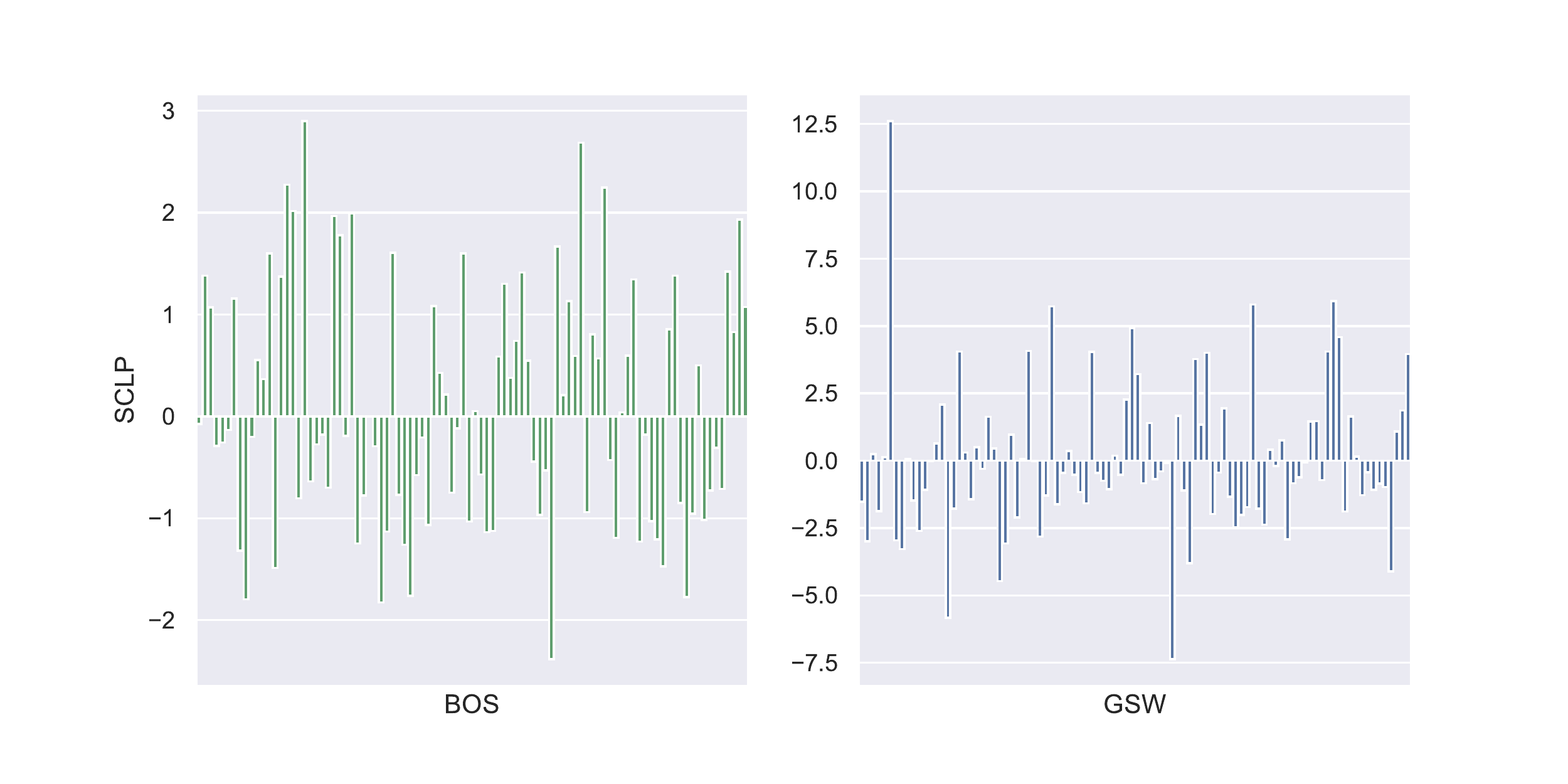}
\caption{Bar graph of third order spectral contributions per log possession (SCLP)
 for BOS and GSW for groups with more than 150 possessions.}
\label{GSWBOS3Bar}
\end{figure}

Table \ref{SmallBOSThirdTable} shows top and bottom three third-order effects for the 15-16 Boston Celtics. (The appendix includes more complete tables for Boston including effects of all orders.) Figure \ref{GSWBOS3Bar} gives contrasting bar plots of the third-order effects for both Boston and Golden State. 
\begin{table}[hbt]
{\footnotesize
\begin{center}
\begin{tabular}{lllrrr}
\hline 
P1 & P2 &P3 & SCLP & PM & Poss \\ \hline
    Evan Turner &  Kelly Olynyk  &  Jonas Jerebko      &  2.9 & 110.1 &  879\\
  Isaiah Thomas&  Avery Bradley & Jared Sullinger     &   2.7&  177.7 & 2642\\
   Avery Bradley &   Jae Crowder & Jared Sullinger     &   2.3 & 139.3&  2216\\
\hline\hline
   Isaiah Thomas &     Evan Turner  & Kelly Olynyk  &     -1.8& -30.9 & 870\\
  Avery Bradley&  Jared Sullinger & Jonas Jerebko  &     -2.3& -11.7 & 194\\
  Isaiah Thomas &   Avery Bradley & Jonas Jerebko   &    -2.4 & -1.6 & 290\\
\hline
\end{tabular} 
\caption{Top and bottom three third-order effects for BOS with at least 150 possessions.}\label{SmallBOSThirdTable}
\end{center}
}
\end{table} 
The Celtics have fewer highly dominant groups. In particular, we note that the spectral signature of the Celtics is distinctly different from that of the Warriors in that Boston lacks anything resembling the big-three of Golden State. While SCLP values are not directly comparable across teams (they depend, for instance, on the norm of the overall team success function when projected into each effect space), the relative values within an effect-space are comparable. Similarly, the SCLP values also depend on the norm of the interpretable vector used in Mallow's method. As a result, the values are not directly comparable across effect spaces-- a problem we return to below.


In fourth and fifth-order spaces the numbers of high-possession groups begins to decline, as alluded to above. (See appendix for complete tables.)  Still, it is interesting to note that spectral analysis flags the Warriors small lineup of Green--Curry--Thompson--Barnes--Iguodala as the team's best, even over the starting lineup with Bogut replacing Barnes. It also prefers two lesser-used lineups to the Warriors' second most-used lineup of Green--Curry--Thompson--Bogut--Rush.  Also of note is the fact that Golden State's best group of three and best group of four are both subsets of the starting lineup-- another instance of stacking of positive effects--while neither of Boston's best groups of three or four are part of their starting lineup. 

\section{Connection With Linear Models}
\label{sec:LM Section}
Before moving on, we consider the connection between spectral analysis and a related approach via linear regression which will likely be more familiar to the sports analytics community. 

Recalling our assumption of a 15 man roster, consider the problem of modeling a lineup's plus-minus, given by $f(L)$ for lineup $L$, using indicator variables that correspond to all possible groups of players. Label the predictor variables $X_1$, $X_2$,\ldots $X_p$, where each variable corresponds to a group of players (with some fixed group order). Thus, the variable $X_i$ is 1 when the players from group $i$ are on the floor, and zero otherwise. If the first fifteen variables are the indicator functions of the individual players $X_1, X_2,\ldots X_{15}$, then the group variables, the $X_i$ for $i>15$, are interaction terms. For instance, the variable corresponding to the group $\{1,2,3\}$ is $X_1X_2X_3$. This approach is therefore similar to an adjusted plus minus with interactions approach. Including all possible group effects, however, means that the number of predictors is quite large and depending on the number of observations, we may be in a situation where $p>>N$. Moreover, the nature of player usage in lineups means that there is a significant multicollinearity issue. Consequently, an attempt to quantify group effects in a regression model of this sort will rely on a shrinkage technique like ridge regression. 

Let $N$ be the number of lineups, and $y=f(L)$, an $N\times 1$ column vector.  Let $\bf X$ be the $N\times (p+1)$ matrix whose first column is the vector of all ones and where the $i$-th row consists of the binary value of each predictor variable for the $i$-th player group. The vector of ridge coefficients $\hat{\beta}^{\text{ridge}}$ minimizes the penalized residual sum of squares: $\argmin_\beta \left\{ \| y-{\bf X}\beta \|^2 +\lambda\sum_{i=1}^p\beta_i^2 \right\}$. The non-negative parameter $\lambda$ serves as a penalty on the $L_2$-norm of the solution vector. (The intercept is not included in the ridge penalty.)
The ridge approach reduces the variability exhibited by the least squares coefficients in the presence of multicollinearity by shrinking the coefficient estimates in the model towards zero (and toward each other). One can show that ridge regression uses the singular values of the covariance matrix associated with the centered version of ${\bf X}$ to disproportionately shrink coefficients associated with inputs where the data exhibits lower degrees of variance. See \cite{friedman2001elements} for details.

The fitted coefficients $\hat{\beta_0},\hat{\beta}_1,\ldots\hat{\beta}_p$ in the ridge regression model 
attempt to measure the contribution of group $i$ while controlling for the contributions of all other groups and individuals. We note that this modeling approach resembles work in \cite{Sill:2010}, \cite{grassetti2019estimation}, and \cite{grassetti2019play}, though there are key differences which we explore below. In particular, note that we model group contributions aggregated over all opponents, and without controlling for the quality of the opponents faced. This simplified approach allows for a more direct comparison with the results of spectral analysis above.   

Tables \ref{RidgeIndsPairs} and \ref{RidgeTriples} give the ridge regression coefficients associated with the top 5 individuals, pairs, and triples for the Warriors.  
\begin{table}[h!]
{\footnotesize
\begin{center}
\begin{tabular}{lcc|cllc}
\hline 
Individual & Estimate &\  & \ & P1 & P2 & Pair Estimate \\
\hline
Draymond Green  &0.28&\  &\ &  Draymond Green&Stephen Curry&0.65 \\
Stephen Curry  &0.25&\  &\ & Stephen Curry&Andrew Bogut&0.53 \\
Klay Thompson& 0.15&\ &\ & Stephen Curry&Klay Thompson&0.47 \\
Andrew Bogut & 0.14&\ &\ & Draymond Green&Klay Thompson&0.47 \\
Festus Ezeli  & 0.02&\ &\ & Draymond Green&Andrew Bogut&0.46 \\
\hline
\end{tabular}
\caption{Best individuals and pairs using the linear model.}\label{RidgeIndsPairs}
\end{center}
}
\end{table}
\begin{table}[hbt]
{\footnotesize
\begin{center}
\begin{tabular}{lllr}
\hline 
P1 & P2 &P3 & Estimate  \\ \hline
Draymond Green&Stephen Curry&Andrew Bogut&      1.61\\
Stephen Curry&Klay Thompson&Andrew Bogut&      1.49\\
Draymond Green&Stephen Curry&Klay Thompson&      1.39\\
Draymond Green&Klay Thompson&Andrew Bogut&      1.24\\
Draymond Green&Klay Thompson&Harrison Barnes&      1.03\\
\hline
\end{tabular} 
\caption{Top triples according to the linear model.}\label{RidgeTriples}
\end{center}
}
\end{table} 
Comparing with Tables \ref{GSWFirstTable}, \ref{GSWSecondTable}, and \ref{GSWThirdTable} shows both some overlap in the top rated groups, but also significant differences with respect to both ordering and magnitude of contribution. In particular, the linear model appears to value the contributions of Andrew Bogut considerably more than spectral analysis. It is also notable that spectral analysis identifies a clearly dominant big three of Green--Curry--Thompson, in contrast to the considerably different result arising from the modeling approach which ranks that group third.





We can interpret the linear model   determined by $\hat{\beta}^{\text{ridge}}$ as giving a similar decomposition to the spectral decomposition in $(\ref{decomposition})$. For each lineup $L$ we have predicted success given by
\begin{equation}
\label{lm}
\hat{\bf y} = {\bf X}_L\hat{\beta}^{\text{ridge}}
\end{equation}
where ${\bf X}_L$ is now the ${15 \choose 5} \times (p+1)$ matrix whose first column is all 1s, and whose $i,j+1$ entry is 1 if the $j$-th player group is part if the $i$-th lineup. (We have fixed a particular ordering of lineups.) The columns of ${\bf X}_L$ (the $X_i$) that correspond to individual players can be understood as spanning a subspace $W_1$ analogous to $V_1$ in (\ref{decomposition}). Similarly, $W_2$ is spanned by the columns of ${\bf X}_L$ corresponding to pair interactions, and so on for all groups through full five player lineups. The particular linear combinations in each $W_i$ determined by the respective coordinates of $\hat{\beta}^{\text{ridge}}$ are analogous to the ${\bf pr}_{V_i}f$. In fact, the space of all lineup functions can be written
\begin{equation}
\label{lmdecomp}
V=W_0+W_1+W_2+W_3+W_4+W_5,
\end{equation}
where $W_i$ is the space of interaction effects for groups of size $i$.

Still, there are important differences between (\ref{decomposition}) and (\ref{lmdecomp}). While $V_0$ and $W_0$ are both one-dimensional, for $i\ge 1$ the dimensions of the $W_i$ are strictly larger than those of their $V_i$ counterparts. For instance, $W_5$ includes a vector for each possible set of five players from the original fifteen. Similarly $W_4$ and groups of four, and so on. Thus, the dimension of $W_5$ is 3003 (the number of lineups), which is the same as the dimension of $V$ itself. By contrast the dimension of $V_5$ in (\ref{decomposition}) is only 1638. Similarly the dimension of $W_4$ is 1365 while that of $V_4$ is $350$. Clearly, the decomposition in (\ref{lm}) is highly non-orthogonal (explaining the $+$ rather than $\oplus$ notation). It is easy to find vectors in $W_i$ that overlap with $W_j$ in the sense that their inner product is non-zero. In the context of basketball, the contribution of a group of, for example, $5$ players is not necessarily separate from a constituent group of four (or any other number of) players despite the use of shrinkage methods.

The decomposition in ($\ref{decomposition}$) is special in that it gives minimal subspaces that are invariant under relabeling and mutually orthogonal as described in section \ref{methodology}. As we've seen, spectral analysis achieves this at the expense of easy interpretation of group contributions. This is a drawback to spectral analysis that (\ref{lm}) does not have, and is an appealing feature of regression models. The interaction term associated with a group of $i$ players in a regression model is easy to understand. Still, as we see above one must balance either ease of interpretation, or orthogonality of effects.  

\section{Stability}
\label{sec: Stability}

In this section we take a first step to addressing questions of the stability of spectral analysis. We seek evidence that spectral analysis is indicative of a true signal, and that should the data have turned out slightly differently, the analysis would not change dramatically. Since spectral analysis works on the lineup function $f(L)$, which is aggregated over all of a team's plays involving $L$, we need to introduce variability into the values of $f(L)$. A fully aggregated NBA season is, in a sense, a complete record of all events and lineup outcomes in that season. Still, it seems reasonable to leverage the variability inherent in the many observed results of a lineup's plays, as well as the substitution patterns of coaches, and suggest a bootstrapping approach.      

To that end, we start with the actual 15-16 season for the Boston Celtics.  We can then build a bootstrapped season by sampling plays, with replacement, from the set of all plays in the actual season.  (We sample the same number of plays as in the actual season.) A play is defined as a connected sequence of events surrounding a possession in the team's play-by-play data. For example, a play might involve a sequence like a missed shot, offensive rebound, and a made jump shot; or, a defensive rebound followed by a bad pass turnover. When sampling from a team's plays, a particular lineup will be selected with a probability proportional to the number of plays in which that lineup participated. We generate 500 bootstrapped seasons, process each using the methodology of sections \ref{Data} and \ref{methodology} to produce success functions $f_{\text{boot}}$, and then apply spectral analysis to each. We thus have a bootstrapped distribution of lineup plus-minus and possession values over each lineup $L$, which in turn gives plus-minus and possession distributions of all player-groups. While the the number of possessions played is highly stable for both full-lineups and smaller player-groups, there is considerable variability in plus-minus values over the bootstrapped seasons. Lineups with a significant number of possessions exhibit both positive and negative performance, and the balance between the positive and negative plays is delicate. 

\begin{figure}[htbp]
\centering
\includegraphics[width = \textwidth]{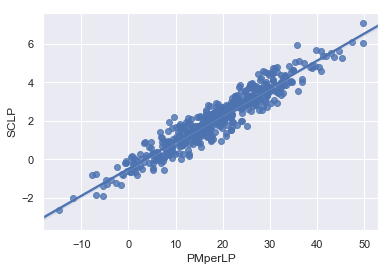}
\caption{Spectral contribution per log possession (SCLP) versus plus-minus per log possession (PMperLP) for Thomas--Bradley--Crowder triple in 500 bootstrapped seasons. Each bootstrapped season consists of sampling plays (connected sequences of game events) with replacement from the set of all season plays. Resampled season data is then processed as in section \ref{Data} and group contributions are computed via spectral analysis as in section \ref{methodology}.}
\label{BOS3BootGroup0}
\end{figure}

The variability in group PM presents a challenge in gauging the stability of the spectral analysis associated with a player group. Take, for example, the Thomas--Bradley--Crowder triple for the Celtics.  The actual season's plus-minus for this group was 
154.8 in 2572 possessions. Over the bootstrapped seasons the group has means of 145.9 and 2574.1 for plus-minus and possessions, respectively. On the other hand, the standard deviation of the plus-minus values is 82.8 versus only 47.7 for possessions. Thus, some of the variability in the spectral contribution of the group over the bootstrapped seasons should be expected since, in fact, the group was less effective in some of those seasons. Figure \ref{BOS3BootGroup0} shows SCLP plotted against PMperLP for the Thomas--Bradley--Crowder triple in 500 bootstrapped seasons. Of course, spectral analysis purports to do more than raw plus-minus by removing otherwise confounding colinearities and overlapping effects. Not surprisingly, therefore, we still see variability in SCLP within a band of plus-minus values, but the overall positive correlation, whereby SCLP increases in seasons where the group tended to outscore its opponents, is reasonable. 


Also intuitively, the strength of the correlation between group plus-minus and spectral contribution depends on the number of possessions played. Fewer possessions means that group's contribution is more dependent on other groups and hence exhibits more variability. The mean possessions for the Thomas--Bradley--Crowder triple in Fig.\ref{BOS3BootGroup0} is 2574, and has a Pearson correlation of $r=0.953$. The group Thomas--Turner--Zeller, on the other hand, has $r=0.688$ with a mean of 305 possessions.
A group like Jared Sullinger--Marcus Smart is particularly interesting. 
This pair has a season plus-minus of 25.0 in 1116 possessions. In 500 bootstrap seasons, they have a mean plus-minus of 23.6 and mean possessions of 1118.3. The value of the group's plus-minus is negative in only $32.4\%$ of those seasons. Should this group, therefore, be considered effective overall? Spectral analysis answers with a fairly emphatic {\it no}. After removing other group contributions their SCLP as a pure pair is negative in $90.6\%$ of bootstrapped seasons, while still exhibiting strong correlation with overall plus-minus ($r=0.73$).
Similarly, the Bradley--Smart pair has a season plus-minus of 45.3 in 1679 possessions
In 500 bootstrap seasons, they have a mean plus-minus of 40.4 and mean possessions of 1679. Their plus-minus is negative in $27\%$ of those seasons while their spectral contribution is negative in $81\%$ of bootstrapped seasons.


\section{Importance of Effect Spaces}
\label{sec: importance}

Another natural question is how to value the relative importance of the group-effect spaces. One way to gauge importance uses the squared $L_2$ norm of the success function in each space.  Since the spaces are mutually orthogonal, we have $\|f\|^2=\|f_1\|^2+\|f_2\|^2+\|f_3\|^2+\|f_4\|^2+\|f_5\|^2$. (Recall that $f_i$ is the projection of $f$ onto the $i$-th order effect space $V_i$.) One can then measure the total mass of $f$ that is concentrated in each effect space. For example, if we found that the mass of the success function was concentrated in the mean space, and thus, a constant function gave a good approximation to $f$, we could conclude that the particular lineup used by this team was largely irrelevant-- the success of the team never strayed far from the mean and was not strongly affected by any groups. This would be an easy team to coach. Of course, this is not the case in basketball, as evidenced by the $L_2$ norm squared distribution of the sample of teams in Table \ref{L2 Table}. 

\begin{table}[htbp]
  \centering
      \begin{tabular}{l|rrrrrr} 
      \hline 
    Team & $V_0$   & $V_1$  &  $V_2$    & $V_3$  & $V_4$   & $V_5$  \\ \hline
   BOS   &  0.001 &   0.012	  & 0.048 &	 0.138 & 	0.297 & 0.504 \\
    CLE  & 0.003 & 0.021 & 0.058 & 0.150  & 0.301 & 0.467 \\
    GSW & 0.003 & 0.031 & 0.092 & 0.203 & 0.312 & 0.360 \\
    HOU  & 0.000 & 0.007 & 0.037 & 0.123 & 0.285 & 0.548 \\
    OKC   & 0.001 & 0.011 & 0.038 & 0.137 & 0.304 & 0.510 \\
    POR   & 0.000 & 0.004 & 0.027 & 0.112 & 0.289 & 0.568 \\
    SAS   & 0.007 & 0.027 & 0.072 & 0.173 & 0.294 & 0.427 \\ \hline
    Null  & 0.000 & 0.005 & 0.03  & 0.117 & 0.303 & 0.545 \\
    \end{tabular}%
    \caption{Distribution of the squared $L_2$-norm of the team success function over the effect spaces.}\label{L2 Table}
\end{table}%
 
By this measure, the higher-order spaces are dominant as they hold most of the mass of the success function.  An issue with this metric, however, is the disparity in the dimensions of the spaces. Because $V_5$ is 1638-dimensional, we might expect the mass of $f$ to be disproportionately concentrated in that space. In fact, a random unit vector projected into each of the effect spaces would be, on average, distributed according to the null distribution in Table \ref{L2 Table}, with mass proportional to the dimension of each of the spaces in question. 

Moreover, we can take the true success function of a team and break the dependence on the actual player groups as follows. Recall that the raw data $f$ records the plus-minus for each of the possible 3003 lineups. We then take $f$ and randomly permute the values so that there is no connection between the lineup and the value associated with that lineup. Still, however, the overall plus-minus and mean of $f$ are preserved. We can then run spectral analysis on the permuted $f$ and record the distribution of the squared $L_2$ norm in each space. Repeating this experiment 500 times for both GSW and BOS give means in Table \ref{gswbosnull} that exactly conform to the null distribution in Table \ref{L2 Table}.    

\begin{table}
\centering
    \begin{tabular}{l|rr} 
    \hline
 Space & BOS & GSW\\ 
 \hline
First &     0.005    &   0.005\\
Second &    0.030  &   0.030\\
Third &     0.117   &   0.116\\
Fourth &    0.302   &   0.302\\
Fifth&      0.543  &    0.544\\
\hline
\end{tabular}
\caption{Average fraction of squared $L_2$ mass by order effect space using randomly permuted success function.}\label{gswbosnull}
\end{table}%

An alternative measure of the importance of each effect space is given by measuring the extent to which projections onto $V_i$ deviate from the null distribution.  By this measure of importance, there is some preliminary evidence that strong teams shift the mass of $f$ from $V_5$ into lower-order spaces, particularly $V_1$, $V_2$, and $V_3$. This is interesting as it agrees with the idea that building an elite team requires a group of three stars. Using all 30 NBA teams, we compute correlations of $r=0.51$, $r=0.58$ and $r=0.55$, respectively, between win-percentage and the projected mass $f$ in the first, second, and third-order spaces. Win-percentage and fifth-order projection have correlation coefficient $r=-0.54$. As pointed out in \cite{Diaconis:1989}, however, care must be taken when looking at deviation from the null distribution if the projections are highly structured and lie close to a few of the interpretable vectors. This is a direction for further inquiry.

\section{Conclusion}

Spectral analysis proposes a new approach to understanding and quantifying group effects in basketball.  By thinking of the success of a team as function on lineups, we can exploit the structure of functions on permutations to decompose the team success function. The resulting Fourier expansion is naturally interpreted as quantifying the group effects to overall team success.  The resulting analysis brings insight into important and difficult questions like which groups of players work effectively together, and which do not. Furthermore, the spectral analysis approach is unique in addressing questions of lineup synergies by presenting an EDA summary of the actual team data without making the kind of modeling or skill-based assumptions of other methods.  

There are several directions for future work. First, the analysis presented used raw lineup level plus-minus to measure success. This approach has the advantage of keeping the analysis tethered to data that is intuitive, and helps avoid pitfalls arising from low-possession lineups. Still, adjusting the lineup level plus-minus to account for quality of opponent, for example, seems like a valuable next step. Another straight forward adjustment to raw plus-minus data would involve devaluing so-called garbage time possessions when the outcome of the game is not in question.  

As presented here, spectral analysis provides an in-depth exploratory analysis of a team's lineups. Still, the results of spectral analysis could also add valuable inputs to more traditional predictive models or machine learning approaches to projecting group effects. Similarly, it would be interesting to use spectral analysis as a practical tool for lineup suggestions. While the orthogonality of the spectral decomposition facilitates valuation of pure player-groups, the question of lineup construction realistically begins at the level of individuals and works up, hopefully stacking the contributions of individuals with strong pairs, triples, and so-on. A strong group of three, for instance, without any strong individual players may be interesting from an internal development perspective, or at the edges of personnel utility, but may also be of limited practical value from the perspective of constructing a strong lineup. Development of a practical tool  would likely require further analysis of the ideas in sections \ref{sec: Stability} and \ref{sec: importance} based on ideas in \cite{Diaconis:1998}. For example, given data (a function on lineups), we might fix the projection of that data onto certain spaces (like the first or second order), and then generate new sample data conditional on that fixed projection. The resulting projections in the higher-order spaces would give some evidence for how the fixed lower-order projections affect the mass of $f$ in the higher-order effects spaces. This would help give a more detailed sense of variability of projections, and a more definitive answer to the question of which spaces are most important, and how the spectral signature of a team correlates with team success. With that information in place, however, one can build tools to suggest lineup replacements that maximize the stacking of a team's most important groups.

\bibliographystyle{plainnat}
\bibliography{BBSpectralAnalysis}

\pagebreak

\section{Appendix}
We include more detailed tables reporting group effects of all orders for both Golden State and Boston.

\begin{table}[hb]
\begin{center}
\begin{tabular}{lccc}
\hline 
Player & SCLP  & PM & Poss\\
\hline
Draymond Green&       17.2&  1038.4 & 5800\\
Stephen Curry   &    15.9 &  978.7&  5610\\
Klay Thompson    &   12.0 &  808.6&  5453\\
Andre Iguodala     &   3.5 &  436.1 & 3516\\
Andrew Bogut    &    2.8 &  403.6 & 2951\\
Harrison Barnes   &     2.2&   384.7 & 4138\\
Festus Ezeli    &   -1.9&   225.0&  1550\\
Shaun Livingston &      -2.0  & 211.1&  2980\\
Leandro Barbosa  &     -5.8 &   70.6 & 2144\\
Brandon Rush     &  -7.1 &   23.3 & 2087\\
Marreese Speights &      -7.4 &   20.0&  1630\\
Ian Clark   &    -9.8 &  -51.9&  1108\\
Anderson Varejao &     -11.1 &  -34.4&   368\\
Jason Thompson    &  -11.2&   -33.8 &  339\\
James Michael McAdoo &     -12.1&   -85.0&   526\\
\hline
\end{tabular}
\caption{All first order effects for GSW.}\label{AGSWFirstTable}
\end{center}
\end{table}

\begin{table}
{\footnotesize
\begin{center}
\begin{tabular}{llccc}
\hline 
P1 &P2 & SCLP  & PM & Poss\\
\hline
Draymond Green &  Stephen Curry    &   13.3&  979.9 & 5102\\
Stephen Curry      &   Klay Thompson  &     11.2&  827.8 & 4311\\
Draymond Green  &  Klay Thompson   &    11.1&  847.8 & 4678\\
Leandro Barbosa  &  Marreese Speights&        5.3&   76.2  & 983\\
Draymond Green  &  Andre Iguodala&        4.3&  490.0 & 2165\\
Klay Thompson  &  Andre Iguodala  &      4.2 & 411.4 & 1764\\
Stephen Curry    & Andre Iguodala    &    3.9 & 460.0  &2185\\
Klay Thompson  &  Harrison Barnes   &     3.9&  396.0 & 3058\\
Klay Thompson  &   Andrew Bogut    &    3.7&  394.3  &2637\\
Leandro Barbosa  &      Ian Clark      &  3.5 &  -9.6 &  325\\
Draymond Green  &  Harrison Barnes&        3.3&  445.1 & 2634\\
Stephen Curry      &  Harrison Barnes &       3.2&  423.3 & 2809\\
Marreese Speights   &     Ian Clark  &      2.5&  -44.2 &  493\\
Harrison Barnes    &  Andrew Bogut &       2.1&  206.2 & 1527\\
Leandro Barbosa   &  Brandon Rush &       1.8&  -22.4 &  638\\
Brandon Rush      &  Ian Clark      &  1.7 & -64.6&   463.0\\
Andre Iguodala    &  Festus Ezeli   &     1.7  &152.6&   999.0\\
Stephen Curry    & Andrew Bogut    &    1.6 & 378.5&  2530.0\\
Shaun Livingston &   Marreese Speights &       1.6&   17.8 & 1014.0\\
Leandro Barbosa &    Festus Ezeli    &    1.3 &  26.2 &  468.0\\ 
\hline\hline
Stephen Curry         &           Brandon Rush   &    -2.5&  140.7 & 1260.0 \\                 
Harrison Barnes      &           Marreese Speights&       -2.9 & -48.8 &  794.0   \\             
Draymond Green    &            Brandon Rush      & -3.0 & 144.4 & 1266.0\\                       
Klay Thompson      &              Festus Ezeli      & -3.1 & 138.9  & 824.0 \\                   
Harrison Barnes     &             Brandon Rush     &  -3.1 & -50.1&   546.0   \\                 
Harrison Barnes     &             Festus Ezeli     &  -3.2  &  9.5  & 598\\                    
Draymond Green   &             Leandro Barbosa&       -3.3 & 154.7 &  860 \\               
Klay Thompson     &             Shaun Livingston   &    -3.6 & 111.8 & 1412  \\                
Stephen Curry       &             Leandro Barbosa     &  -3.7 & 126.3  & 883   \\              
Draymond Green   &            Marreese Speights    &   -4.0 & 129.7 &  492 \\                  
Andre Iguodala      &             Brandon Rush    &   -4.4&  -59.9&   399  \\                  
Klay Thompson      &                 Ian Clark    &   -5.1 &  10.3 &  498  \\                  
Stephen Curry       &            Marreese Speights &      -5.7 &  73.5   &423 \\               
Klay Thompson      &           Marreese Speights   &    -6.4  & -5.1 &  581  \\                
Klay Thompson      &            James Michael McAdoo &      -7.1 & -28.9&   241 \\             
Draymond Green   &             Ian Clark    &   -7.2 &  33.3  & 424  \\                        
Klay Thompson      &             Leandro Barbosa &      -7.2 &   4.8   &349 \\                 
Stephen Curry        &             Ian Clark    &   -8.1  & 14.0 &  220  \\                    
Draymond Green   &            Anderson Varejao &      -9.5  &  7.2  & 217\\                    
Stephen Curry        &            Anderson Varejao  &    -10.1 & -26.9  & 237 \\  \hline     
\end{tabular}
\caption{Second order effects for GSW with at least 200 possessions.}\label{AGSWSecondTable}
\end{center}
}
\end{table}

\begin{table}
{\footnotesize
\begin{center}
\begin{tabular}{lllrrr}
\hline 
P1 & P2 &P3 & SCLP & PM & Poss \\ \hline
Draymond Green &   Stephen Curry  &    Klay Thompson    &   12.6 & 812.7   &    4085  \\
Draymond Green   & Klay Thompson  & Harrison Barnes  &      5.9 & 427.3 &  2473 \\
Draymond Green  &  Stephen Curry  &   Andre Iguodala    &    5.8 & 464.8  & 1830 \\
Stephen Curry  &  Klay Thompson   & Harrison Barnes   &     5.7 & 416.5  &  2431 \\
Stephen Curry &   Klay Thompson   &    Andrew Bogut    &    4.9&  382.2 &   2296 \\
Stephen Curry&  Leandro Barbosa & Marreese Speights  &      4.6 &  84.6   &  245 \\
Draymond Green &   Stephen Curry    &   Andrew Bogut     &   4.1 & 377.4   & 2346\\
Draymond Green   & Stephen Curry  &  Harrison Barnes  &      4.1 & 411.3   &2421 \\
Stephen Curry &  Andre Iguodala    &   Festus Ezeli    &    4.1 & 197.2  &   633 \\
Draymond Green &   Klay Thompson   &  Andre Iguodala &       4.0&  388.4 &  1418\\ 
Draymond Green   & Klay Thompson &      Andrew Bogut &       4.0 & 359.8  & 2409 \\
Stephen Curry &   Klay Thompson   &  Andre Iguodala   &     4.0 & 377.0 &   1270 \\
Draymond Green & Leandro Barbosa & Marreese Speights &       3.8 &  88.9 &   248 \\
Draymond Green&   Andre Iguodala  &     Festus Ezeli    &    3.2 & 180.8  &  569 \\
Klay Thompson&  Harrison Barnes &    Andre Iguodala    &    2.3 & 199.5  &   671 \\ \hline\hline
Leandro Barbosa & Marreese Speights & Ian Clark & -2.1  & -0.2  & 203 \\
Draymond Green & Andre Iguodala & Andrew Bogut & -2.4  & 79.4  & 535 \\
Draymond Green & Shaun Livingston & Andrew Bogut & -2.5  & 47.4  & 370 \\
Klay Thompson & Harrison Barnes & Marreese Speights & -2.6  & -31.4 & 323 \\
Stephen Curry & Andre Iguodala & Andrew Bogut & -2.8  & 70.2  & 541 \\
Klay Thompson & Harrison Barnes & Festus Ezeli & -2.9  & -1.2  & 353 \\
Draymond Green & Stephen Curry & Leandro Barbosa & -3.0  & 126.9 & 687 \\
Stephen Curry & Klay Thompson & Shaun Livingston & -3.0    & 121.3 & 530 \\
Klay Thompson & Harrison Barnes & Brandon Rush & -3.1  & -1.0    & 265 \\
Draymond Green & Harrison Barnes & Festus Ezeli & -3.3  & 16.0    & 326 \\
Stephen Curry & Andre Iguodala & Brandon Rush & -3.8  & -13.5 & 207 \\
Draymond Green & Stephen Curry & Marreese Speights & -4.1  & 97.9  & 299 \\
Draymond Green & Klay Thompson & Marreese Speights & -4.5  & 52.2  & 250 \\
Draymond Green & Klay Thompson & Ian Clark & -5.8  & 9.8   & 316 \\
Draymond Green & Stephen Curry & Ian Clark & -7.4  & 14.5  & 205 \\
\hline
\end{tabular} 
\caption{Third order effects for GSW with at least 200 possessions.}\label{AGSWThirdTable}
\end{center}
}
\end{table}

\begin{table}[h!]
{\footnotesize
\begin{center}
\begin{tabular}{llllccc}
\hline
P1&P2&P3&P4&SCPLP&PM&Poss\\
\hline 
Draymond Green &  Stephen Curry &     Klay Thompson &    Harrison Barnes   & 8.7  & 401.6&2271  \\
Draymond Green&    Stephen Curry&      Klay Thompson  &      Andrew Bogut &   7.8  &365.7&2159 \\
Draymond Green &   Stephen Curry &     Klay Thompson&      Andre Iguodala  &  7.7  &364.9&  1157\\
Draymond Green &  Andre Iguodala  &Shaun Livingston &       Festus Ezeli    &   3.9  & 76.8& 201 \\
Draymond Green &   Stephen Curry &   Leandro Barbosa &  Marreese Speights  &  3.9  & 67.8& 173\\
Draymond Green &   Stephen Curry  &   Andre Iguodala &       Festus Ezeli &   3.8  &170.3& 526\\
Stephen Curry &   Klay Thompson &   Harrison Barnes &     Andre Iguodala    & 3.3  &171.3 &451\\
Draymond Green  &  Stephen Curry&    Harrison Barnes &       Andrew Bogut  &  2.8 & 162.3 & 1165\\
Stephen Curry &  Andre Iguodala&   Shaun Livingston &       Festus Ezeli    & 2.7  & 64.9& 201\\ 
Draymond Green &   Stephen Curry &     Klay Thompson  &      Brandon Rush &   2.4 & 177.9&870\\
Draymond Green &   Stephen Curry &   Harrison Barnes&      Andre Iguodala  &  2.3  &157.7& 419\\
Draymond Green &   Klay Thompson &   Harrison Barnes &     Andre Iguodala &   2.0 & 158.0&417\\
Draymond Green &   Klay Thompson &   Harrison Barnes  &      Andrew Bogut &   1.8 & 158.4 & 1221\\
Draymond Green&    Stephen Curry &  Shaun Livingston  &      Festus Ezeli   & 1.7  & 75.7& 198\\
Stephen Curry&    Klay Thompson  &  Harrison Barnes    &    Andrew Bogut   & 1.5  &154.0& 1235\\ 
\hline
Draymond Green&   Stephen Curry &   Andre Iguodala   &   Andrew Bogut&  -1.3       & 79.3 &433\\
Draymond Green&   Stephen Curry&   Harrison Barnes & Shaun Livingston&  -1.7     &57.4 &214\\
Harrison Barnes & Andre Iguodala&  Shaun Livingston &     Festus Ezeli &-2.0           &-20.7  &160\\
Draymond Green &  Stephen Curry&     Klay Thompson&  Shaun Livingston&  -2.1   & 116.5  & 485\\
Draymond Green &  Stephen Curry&   Harrison Barnes  &    Brandon Rush&  -2.1     & 23.0  &160\\
Draymond Green &  Klay Thompson&   Harrison Barnes&      Festus Ezeli&  -2.3       & 17.1  &299\\
Stephen Curry &  Klay Thompson &  Harrison Barnes&      Brandon Rush&   -2.4        & 30.3  &152\\
Stephen Curry &  Klay Thompson &  Harrison Barnes &     Festus Ezeli&   -2.5          & 19.0  &309\\
Draymond Green&   Stephen Curry&   Harrison Barnes&      Festus Ezeli&  -3.0        &17.7  &309\\
Draymond Green&   Klay Thompson &   Andre Iguodala&  Shaun Livingston&  -3.0  &18.2 &261\\
\hline
\end{tabular} 
\caption{Fourth order effects for GSW with at least 150 possessions.}\label{AGSWFourthTable}
\end{center}
}
\end{table}

\begin{table}[h!]
{\scriptsize
\begin{center}
\begin{tabular}{lllllccc}
\hline 
P1 &P2 & P3&P4&P5&SCLP  & PM &Poss\\
\hline
Draymond Green &    Stephen Curry &     Klay Thompson&    Harrison Barnes  &  Andre Iguodala &      10.3 &      152.9  &      372    \\  
Draymond Green &     Stephen Curry &     Klay Thompson&    Harrison Barnes &  Andrew Bogut&             7.0  &     142.0&       1140        \\
Draymond Green &     Stephen Curry &    Andre Iguodala&   Shaun Livingston &  Festus Ezeli&             6.9  &      67.3  &      160       \\
Draymond Green &     Stephen Curry&      Klay Thompson&       Andrew Bogut&   Brandon Rush&             5.2  &      88.2  &      535        \\
Draymond Green &     Stephen Curry&      Klay Thompson &    Andre Iguodala&   Andrew Bogut&             5.1 &       98.2 &       310       \\
Draymond Green &     Stephen Curry&      Klay Thompson&     Andre Iguodala&   Festus Ezeli&             4.9  &      85.7 &       266        \\
Draymond Green &     Stephen Curry&      Klay Thompson&   Shaun Livingston &  Andrew Bogut       &      1.9  &      42.7 &       112   \\    
Harrison Barnes&   Shaun Livingston&    Leandro Barbosa&       Brandon Rush&  Marreese Speights&             1.8   &      6.4 &        87 \\  
Draymond Green&      Stephen Curry &     Klay Thompson &   Harrison Barnes&   Shaun Livingston       &      0.7 &       42.7 &       175   \\
\hline 
Harrison Barnes &    Andre Iguodala&   Shaun Livingston &   Leandro Barbosa&  Marreese Speights      &      -0.8  &      -3.1 &       172   \\
Harrison Barnes &    Andre Iguodala&   Shaun Livingston&    Leandro Barbosa&  Festus Ezeli      &      -1.3  &      -9.9  &      102       \\
Draymond Green&      Stephen Curry&      Klay Thompson&    Harrison Barnes &  Festus Ezeli      &      -1.9 &       20.1  &      283       \\
Draymond Green &     Stephen Curry&      Klay Thompson&    Harrison Barnes &  Brandon Rush      &      -2.3  &      28.0&        123\\       
Draymond Green &     Stephen Curry&      Klay Thompson  &  Andre Iguodala &    Shaun Livingston      &      -5.1 &        2.3 &        98    \\
Draymond Green &     Stephen Curry&      Klay Thompson &   Harrison Barnes&   James Michael  McAdoo    &        -5.9  &     -14.2  &       91\\
\hline
\end{tabular}
\caption{Fifth order effects for GSW with at least 80 possessions.}\label{AGSWFifthTable}
\end{center}
}
\end{table}

\begin{table}[h!]
{\footnotesize
\begin{center}
\begin{tabular}{llccc}
\hline 
P1 &P2 & SCLP  & PM & Poss\\
\hline
Andrew Bogut  &  Klay Thompson& 3.7 &      394.3&             2637\\ 
Andrew Bogut &   Harrison Barnes &   2.1 &   206.2&           1527 \\ 
Andrew Bogut &   Stephen Curry &    1.6 &    378.5&            2530 \\  
Andrew Bogut &   Draymond Green & 0.8 &    371.5&           2596\\   
Andrew Bogut &   Brandon Rush &   0.7  &        54.5&        733\\ 
Andrew Bogut &   Ian Clark &   -0.3  &        6.1&            198 \\ 
Andrew Bogut &   Shaun Livingston &-0.6  &  77.4&             573 \\    
Andrew Bogut &   Leandro Barbosa &   -1.6 &   16.3&            166 \\   
Andrew Bogut  &  Andre Iguodala &   -2.1 &      107.0 &          785 \\ 
\hline
\end{tabular}
\caption{Pairs involving Andrew Bogut (with at least 150 possessions).}\label{BogutPairs}
\end{center}
}
\end{table}

\begin{table}[h!]
{\footnotesize
\begin{center}
\begin{tabular}{llccc}
\hline 
P1 &P2 & SCLP  & PM &Poss\\
\hline
Shaun Livingston   &   Anderson Varejao  &      2.0  & -1.5   & 174    \\
  Shaun Livingston   &  Marreese Speights  &      1.6 &  17.8  & 1014  \\
    Shaun Livingston    &   Draymond Green    &   1.2 & 323.6  & 1486\\
  Shaun Livingston     &        Ian Clark    &    0.9 & -25.7   & 378          \\
  Shaun Livingston     &  Leandro Barbosa   &     0.9 &  15.2  & 1210    \\
  Shaun Livingston & James Michael McAdoo &       0.8 & -41.6   & 180\\
  Shaun Livingston   &       Festus Ezeli   &     0.4  & 49.0  &  654        \\
   Shaun Livingston      &    Stephen Curry &    -0.1 & 265.5   &1120        \\
  Shaun Livingston   &       Andrew Bogut    &   -0.6  & 77.4  &  573        \\
  Shaun Livingston     &    Harrison Barnes  &   -1.1  & 55.2  & 1475         \\
   Shaun Livingston    &  Andre Iguodala      &  -1.3  & 65.2  & 1605          \\
  Shaun Livingston   &       Brandon Rush    &   -1.5 & -63.2   & 536         \\
  Shaun Livingston     &    Klay Thompson   &   -3.6 & 111.8  & 1412        \\
\hline
\end{tabular}
\caption{Pairs involving Shaun Livingston (with at least 150 possessions).}\label{LivingstonPairs}
\end{center}
}
\end{table}

\begin{table}
{\footnotesize
\begin{center}
\begin{tabular}{lllrrr}
\hline 
Player 1 & Player 2 &Player 3 & SCLP & PM & Poss \\ \hline
Klay Thompson&  Harrison Barnes&  Shaun Livingston &      -1.1 &  35.3   &733   \\
Draymond Green&   Andre Iguodala&  Shaun Livingston&       -1.3&   92.5 &  630  \\
Draymond Green &   Klay Thompson&      Festus Ezeli&       -1.4 & 151.4   & 721 \\
Stephen Curry  &  Klay Thompson &     Festus Ezeli &      -1.5&  152.4   & 694 \\
Draymond Green&    Klay Thompson & Shaun Livingston &      -1.6&  160.5 &  929\\  
Draymond Green &   Stephen Curry  &    Brandon Rush&       -1.7&  153.9  &1116 \\
Draymond Green &  Andre Iguodala&      Andrew Bogut &      -2.4&   79.4  &  535 \\ 
Stephen Curry &  Andre Iguodala  &    Andrew Bogut &      -2.8 &  70.2  &  541 \\
Draymond Green&    Stephen Curry &  Leandro Barbosa &      -3.0&  126.9 &  687\\  
 Stephen Curry &   Klay Thompson & Shaun Livingston   &    -3.0&  121.3  & 530  \\
\hline
\end{tabular} 
\caption{Worst triples for GSW with at least 500 possessions.}\label{AGSWWorst500}
\end{center}
}
\end{table}

\begin{table}[ht]
\begin{center}
\begin{tabular}{lccc}
\hline 
Player & SCLP  & PM & Poss\\
\hline
Isaiah Thomas&3.4&236.5&5388\\
Avery Bradley&3.3&228.5&5099\\
Jae Crowder&3.1&219.5&4685\\
Jared Sullinger&3&210.8&3828\\
Amir Johnson&2&172.8&3580\\
Kelly Olynyk&1.7&154.8&2835\\
Marcus Smart&0.9&125.3&3407\\
Evan Turner&-0.2&81.1&4577\\
Jonas Jerebko&-0.4&71.5&2346\\
RJ Hunter&-3.3&-18.3&624\\
Jordan Mickey&-3.6&6&106\\
David Lee&-3.6&-35&945\\
Tyler Zeller&-3.7&-44.3&1442\\
James Young&-4&-31.3&392\\
Terry Rozier&-4.1&-43&616\\
\hline
\end{tabular}
\caption{First order effects for BOS.}\label{BOSFirstTable}
\end{center}
\end{table}

\begin{table}
{\footnotesize
\begin{center}
\begin{tabular}{llccc}
\hline 
P1 &P2 & SCLP  & PM & Poss\\
\hline
Isaiah Thomas &   Avery Bradley  &      3.5 & 229.8&  3564\\
Evan Turner&    Jonas Jerebko   &     3.0 & 109.3 & 1945\\
Marcus Smart&     Kelly Olynyk   &     3.0 & 141.8&  1298\\
Avery Bradley & Jared Sullinger   &     2.7&  191.2&  2969\\
Tyler Zeller  &      RJ Hunter     &   2.6 &   8.5&   261\\
Isaiah Thomas&  Jared Sullinger&        2.5 & 188.6 & 3315\\
Isaiah Thomas  &    Jae Crowder  &      2.3 & 187.0 & 3668\\
Jae Crowder   &  Amir Johnson    &    2.3 & 162.3 & 2594\\
Isaiah Thomas &    Amir Johnson   &     2.1 & 165.8&  3175\\
Kelly Olynyk  &  Jonas Jerebko    &    2.0  & 95.6 & 1030\\
\hline\hline
Isaiah Thomas   & Evan Turner &      -2.2&   1.0&  2462\\
Avery Bradley &  Tyler Zeller  &     -2.5& -38.0 &  674\\
Jared Sullinger & Jonas Jerebko &      -2.5&  -2.4&   386\\
Isaiah Thomas &  Tyler Zeller &      -2.9& -41.5&   455\\
Avery Bradley &  Terry Rozier &      -3.4& -41.9&   160\\ \hline
\end{tabular}
\caption{Top ten and bottom five second order effects for BOS with at least 150 possessions.}\label{BOSSecondTable}
\end{center}
}
\end{table}
 
\begin{table}
{\footnotesize
\begin{center}
\begin{tabular}{lllrrr}
\hline 
P1 & P2 &P3 & SCLP & PM & Poss \\ \hline
    Evan Turner &  Kelly Olynyk  &  Jonas Jerebko      &  2.9 & 110.1 &  879\\
  Isaiah Thomas&  Avery Bradley & Jared Sullinger     &   2.7&  177.7 & 2642\\
   Avery Bradley &   Jae Crowder & Jared Sullinger     &   2.3 & 139.3&  2216\\
  Isaiah Thomas&  Avery Bradley  &    Jae Crowder    &    2.2 & 154.8&  2572\\
  Isaiah Thomas&  Avery Bradley  &   Amir Johnson   &     2.0&  137.5 & 2351\\
    Evan Turner &  Marcus Smart  &  Jonas Jerebko   &     2.0 &  93.7&  1159\\
    Jae Crowder  &  Evan Turner  &  Jonas Jerebko   &     2.0 &  61.2 &  460\\
 Isaiah Thomas  &  Jae Crowder & Jared Sullinger    &    1.9&  140.7 & 2533\\
 Isaiah Thomas &  Marcus Smart &    Kelly Olynyk    &    1.8 &  85.5&   464\\
  Avery Bradley &   Jae Crowder  &   Amir Johnson&        1.7&  107.3 & 1894\\
\hline\hline
Avery Bradley    &  Jae Crowder   & Evan Turner  &     -1.8 & -7.9 & 708\\
   Isaiah Thomas &     Evan Turner  & Tyler Zeller  &     -1.8& -68.4 & 305\\
   Isaiah Thomas &     Evan Turner  & Kelly Olynyk  &     -1.8& -30.9 & 870\\
  Avery Bradley&  Jared Sullinger & Jonas Jerebko  &     -2.3& -11.7 & 194\\
  Isaiah Thomas &   Avery Bradley & Jonas Jerebko   &    -2.4 & -1.6 & 290\\
\hline
\end{tabular} 
\caption{Top ten and bottom five third order effects for BOS with at least 150 possessions.}\label{BOSThirdTable}
\end{center}
}
\end{table} 
 
 \begin{table}
{\footnotesize
\begin{center}
\begin{tabular}{llllrrr}
\hline 
 P1         &    P2  &              P3   &             P4  & SCLP & PM&  Poss\\ \hline
 Avery Bradley  &  Evan Turner  &    Kelly Olynyk &    Jonas Jerebko &     3.1  &  71.8 &   375\\
    Evan Turner &   Marcus Smart  &    Kelly Olynyk &    Jonas Jerebko &   2.7 &   88.0 &   526\\
   Isaiah Thomas&   Avery Bradley   &    Jae Crowder &  Jared Sullinger&    2.6&   120.0 &  2014\\ 
 Isaiah Thomas &  Avery Bradley   &    Evan Turner&   Jared Sullinger &   2.4 &  76.8 &   584\\
Avery Bradley  &   Evan Turner  &    Marcus Smart  &   Jonas Jerebko &   2.1 &   62.8 &   526 \\
 Avery Bradley  &   Jae Crowder&   Jared Sullinger &     Kelly Olynyk &   1.9 &   59.7  &  247\\
Avery Bradley  &  Marcus Smart &     Kelly Olynyk  &   Jonas Jerebko &   1.7  &  55.6 &   304\\
 Isaiah Thomas&   Avery Bradley  &     Jae Crowder &     Kelly Olynyk &   1.7  &  62.8  &  432\\
  Avery Bradley &    Evan Turner &  Jared Sullinger &     Amir Johnson &   1.6 &   42.2 &   343 \\
Avery Bradley  &   Evan Turner  &  Marcus Smart&      Kelly Olynyk  &  1.6  &  44.3  &  423\\ \hline\hline
Jae Crowder   & Evan Turner&  Jared Sullinger & Marcus Smart   &    -1.0  & -21.2 & 180\\ 
   Isaiah Thomas & Avery Bradley   &   Jae Crowder & Marcus Smart &      -1.1 &   2.1 & 281\\   
    Evan Turner &  Marcus Smart   & Jonas Jerebko & Tyler Zeller    &   -1.1   & 1.5 & 408\\
  Isaiah Thomas & Avery Bradley  &   Amir Johnson&  Marcus Smart&       -1.3&  4.6 & 322\\  
 Isaiah Thomas & Avery Bradley &     Evan Turner & Kelly Olynyk&       -2.6  & -24.2 & 225\\ \hline 
\end{tabular} 
\caption{Top ten and bottom five fourth order effects for BOS with at least 150 possessions.}\label{BOSFourthTable}
\end{center}
}
\end{table}

\begin{table}
{\scriptsize
\begin{center}
\begin{tabular}{lllllrrr}
\hline 
P1&P2&P3&P4&P5&SCLP&PM&Poss\\ \hline
  Avery Bradley  &  Evan Turner  &   Marcus Smart  &   Kelly Olynyk &  Jonas Jerebko  &6.1 & 63.0&257  \\
Isaiah Thomas & Avery Bradley &     Jae Crowder & Jared Sullinger &  Kelly Olynyk  & 3.4 & 41.9&202 \\
 Isaiah Thomas & Avery Bradley  &    Jae Crowder & Jared Sullinger&  Amir Johnson &2.9 & 48.8&  1413\\
 Isaiah Thomas&  Avery Bradley &    Evan Turner & Jared Sullinger &  Amir Johnson    &2.4 & 33.8  & 256 \\
 Isaiah Thomas & Avery Bradley &     Jae Crowder  &    Evan Turner  & Jared Sullinger  &1.1 & 23.7 &  148 \\ \hline\hline
Isaiah Thomas & Avery Bradley  &   Jae Crowder   &  Amir Johnson  &  Kelly Olynyk   &-1.2 &  7.3 &  107 \\
  Isaiah Thomas & Avery Bradley & Jared Sullinger  &   Amir Johnson &  Marcus Smart& -1.6  &-3.8&   128\\  
  Isaiah Thomas  &  Jae Crowder  &    Evan Turner & Jared Sullinger &  Amir Johnson & -1.8 & -7.0 &  105 \\ \hline
\end{tabular} 
\caption{Top five and bottom three fifth order effects for BOS with at least 100 possessions.}\label{BOSFifthTable}
\end{center}
}
\end{table}

\end{document}